\documentclass[letterpaper, conference,10pt] {IEEEtran}
\usepackage[ left=0.64in, right=0.64in, top=0.75in, bottom=0.75in]{geometry}
\usepackage{amsmath}
\usepackage{amssymb}
\usepackage{bm}
\usepackage{color}
\usepackage{graphicx}

\usepackage[linesnumbered,ruled,lined]{algorithm2e}

\usepackage[outdir=./]{epstopdf}
\usepackage{caption}
\usepackage{lipsum}
\usepackage{cite}
\usepackage{multirow}
\usepackage{empheq}
\usepackage{fancyhdr}
\usepackage[caption=false, font=footnotesize]{subfig}

\SetKwInput{Input}{input}
\SetKwInput{Output}{output}

\IEEEoverridecommandlockouts
\begin{document}

	\title{ Rate-Splitting Multiple Access for Coordinated Multi-Point Joint Transmission
		\\[-0.2ex]
		}

		\author{
			 
			\IEEEauthorblockN{Yijie Mao$^*$,  Bruno Clerckx$^\dagger$ and Victor O.K. Li$^*$ }
		\IEEEauthorblockA{$^*$The University of Hong Kong, Hong Kong, China,
			$^\dagger$Imperial College London, United Kingdom\\
				Email: $^*$\{maoyijie, vli\}@eee.hku.hk, $^\dagger$b.clerckx@imperial.ac.uk}
			
			 \\[-6.2ex]
			\thanks{This work is partially supported by the U.K. Engineering and Physical
				Sciences Research Council (EPSRC) under grant EP/N015312/1.}}

\maketitle

\thispagestyle{empty}
\pagestyle{empty}

\begin{abstract}
	As a promising downlink multiple access scheme,
	 Rate-Splitting Multiple Access (RSMA) has been shown to achieve superior spectral and energy efficiencies compared with Space-Division Multiple Access (SDMA) and Non-Orthogonal Multiple Access (NOMA) in downlink single-cell systems. By relying on linearly precoded rate-splitting at the transmitter and successive interference cancellation at the receivers, RSMA has the capability of partially decoding the interference and partially treating the interference as noise, and therefore copes with a wide range of user deployments and network loads.  In this work, we further study RSMA in downlink Coordinated Multi-Point (CoMP) Joint Transmission (JT) networks by investigating the optimal beamformer design to maximize the Weighted Sum-Rate (WSR) of all users subject to individual Quality of Service (QoS) rate constraints and  per base station power constraints.  
	Numerical results show that, in CoMP JT, RSMA achieves significant WSR improvement over SDMA and NOMA in a wide range of inter-user and inter-cell channel strength disparities. Specifically, SDMA (resp. NOMA) is more suited to deployments with little (resp. large) inter-user channel strength disparity and large (resp. little) inter-cell channel disparity, while RSMA is suited to any deployment. 
	 We conclude that RSMA provides rate, robustness and QoS enhancements over SDMA and NOMA in CoMP JT networks.
\end{abstract}
\maketitle


%


\section{Introduction}
\par 
The next generation of wireless communication systems requires a paradigm shift so as to meet the increasing demand for high-rate and multimedia data services.
 Among numerous candidate technologies, a novel multiple access scheme, namely  Rate-Splitting Multiple Access (RSMA), has been proposed in \cite{mao2017rate} recently. It has been shown to be more general, robust, and powerful than Space Division Multiple Access (SDMA) and Non-Orthogonal Multiple Access (NOMA).  
Contrary to SDMA that relies on fully treating any residual interference as noise and NOMA that relies on fully decoding interference, RSMA  has the capability of partially decoding the interference and partially treating the interference as noise. It uses linearly precoded Rate-Splitting (RS) at the transmitter to split the messages of users into multiple common and private messages. The messages are jointly transmitted based on a superimposed transmission of common messages (decoded by multiple users) and private messages (decoded by the corresponding users and treated as noise by other users). RSMA adopts Successive Interference Cancellation (SIC)  at the receivers to decode the common messages  before decoding the private messages. 

\par Existing works on RSMA can be classified into two categories. The first category is a single-cell setup as studied in \cite{DoF2013SYang,RS2015bruno,Mingbo2016,RS2016hamdi,mao2017rate,mao2018rate,mao2018EE}. However, those works are limited to the single-cell multi-antenna Broadcast Channel (BC). The co-channel interference caused by Base Stations (BSs) in the adjacent cells are omitted. 
The second category is a multi-cell setup. The RS-assisted multi-cell Interference Channel (IC) has been investigated in 
\cite{TeHan1981,Tse2008,WYu2011rs,chenxi2017brunotopology,chenxi2017bruno}.
Only the recent work \cite{Ahmad2018RScRAN} investigated RS-assisted multi-cell BC by applying RS in Cloud-Radio Access Networks (C-RAN) with partial cooperation among BSs. 
As highlighted in \cite{gesbert2010multicell}, the best means of dealing with the inter-cell interference is the proactive treatment of co-channel interference  via interference-aware multi-cell cooperation, which is also known as  \textit{network} Multiple Input Multiple Output (MIMO) or \textit{Coordinated Multi-Point (CoMP) Joint Transmission (JT)} in the 3rd Generation Partnership Project (3GPP).

A CoMP JT network is first investigated in \cite{networkmimo2001} with the application of Dirty Paper Coding (DPC) and individual power constraint at each single-antenna BS. It was recognized to be identical to the multi-antenna BC with Per-Antenna Power Constraint (PAPC). As shown in \cite{capacityRegion2006HW}, the capacity region of the MIMO BC subject to PAPC is achieved by DPC. The entire DPC region is characterized by establishing the uplink-downlink duality of MIMO BC with PAPC \cite{weiyu2007papc}. The sum rate maximization problem with PAPC in the downlink BC is transformed into  a min-max problem in its dual Multiple Access Channel (MAC) and solved by a barrier interior-point method. A more computationally efficient algorithm to solve the problem is proposed in \cite{pham2017alternating} based on the Alternating Optimization (AO) and Successive Convex Approximation (SCA) algorithms. All these works consider nonlinear DPC, which is computationally prohibitive. It is of practical interest in the multi-cell setup to rely on linear precoding. The design of linear precoder subject to PAPC has been investigated in e.g. \cite{ZF2006Fbocca,weiyu2007papc}. However, the performance of a general framework based on RS (that encompasses SDMA and NOMA) in multi-antenna BC with PAPC and in CoMP JT has never been studied yet.

In this work, we initiate the study of RSMA in CoMP JT  multi-antenna BC where transmit antennas in multiple cells are allowed to fully cooperate with each other and serve multiple users jointly.  
Of particular interest, we ask ourselves how SDMA, NOMA and RSMA compare as a function of the inter-cell and inter-user channel disparities in a CoMP JT network.
To answer this question, for the three multiple access schemes, the precoders from all the BSs are designed together by solving
the Weighted Sum Rate (WSR) maximization problem subject to individual
Quality of Service (QoS) rate requirements and per-BS power constraints.
The problem we investigated is subject to two additional features: 1) per antenna/BS power constraint, 2) inter-cell and inter-user channel disparities.
The performance evaluations show that the rate region of RSMA outperforms that of existing Multi-User Linear Precoding (MU--LP)-based SDMA as well as Superposition Coding with SIC (SC--SIC)-based NOMA in a wide range of inter-user and inter-cell channel strength disparities. 
We show that in a CoMP JT network, SDMA is more suited to the scenarios where there is little inter-user channel strength disparity but large inter-cell channel disparity; NOMA is more suited to the scenarios where there is a large inter-user channel strength disparity but little inter-cell channel disparity; and RSMA always outperforms SDMA and NOMA and is suited to any scenario with any inter-user and inter-cell channel disparity. We also show that RSMA comes much closer to the DPC performance, though it only relies on linear precoding at the transmitter and SIC at the receivers.

%


\vspace{-0.5mm}
\section{System model}
\vspace{-0.5mm}
\label{sec: system model}
\par Consider a downlink multi-cell system of $M$ cells, indexed by $\mathcal{M}=\{1,\ldots,M\}$. In each cell-$m$, there is one single-antenna BS and $K_m$ single-antenna users.  The total number of users is $K\triangleq\sum_{m=1}^{M}K_m$. The users are  indexed by $\mathcal{K}=\{1,\ldots,K\}$.  We are interested in a fully cooperative multi-cell network. The $M$ BSs  work as a 'super BS' and serve $K$ users.

\par Following the generalized RS framework specified in \cite{mao2017rate}, we further extend it to the multi-cell system. The messages $W_1,\ldots,W_K$ intended for  user-$1$ to user-$K$ are jointly processed at the central controller and the optimized transmit signals are sent to the corresponding BSs. Data signals and channel state informations are assumed to be available at the central controller without any delay or imperfection. 

\par The messages of users are split into multiple parts and encoded into different streams. For user-$k$ ($k\in\mathcal{K}$), its intended message $W_k$ is split into multiple sub-messages $\{ W_k^{\mathcal{A}'} | \mathcal{A}' \subseteq \mathcal{K}, k \in \mathcal{A}' \}$. For any subset of users ${\mathcal{A}}$  with user-$k$ included, user-$k$ provides a unique split message $W_k^{\mathcal{A}}$.
The split sub-messages $\{W_{k'}^{\mathcal{A}}|k'\in\mathcal{A}\}$ of all users in the  user set $\mathcal{A}$  are jointly encoded into the stream  $s_{\mathcal{A}}$. It is linearly precoded via the beamforming vector $\mathbf{p}_{\mathcal{A}}\in\mathbb{C}^{M\times 1}$ and transmitted from the BSs to all  users.  $s_{\mathcal{A}}$  is to be decoded by all users in the user set $\mathcal{A}$ and treated as noise by other users. By splitting the message intended for each user into different sub-messages and regrouping the sub-messages, all users enable the capability of dynamic interference management. When user-$k$ decodes $s_{\mathcal{A}}$, it not only decodes its intended sub-message $W_k^{\mathcal{A}}$, but also decodes the interference of other users $\{W_{k'}^{\mathcal{A}}|k'\in\mathcal{A}, k'\neq k\}$.

\par Recall the \textit{$l$-order streams} defined in \cite{mao2017rate} to represent the streams to be decoded by $l$ users. The \textit{stream order} $l$ is defined as the number of users to decode the $l$-order streams. For a given data stream $s_{\mathcal{A}}$ to be decoded by all the users in $\mathcal{A}$, its stream order is $l=|\mathcal{A}|$. Since $\mathcal{A}\subseteq \mathcal{K}$, we have $l\in K$.
For a given $l\in\mathcal{K}$, all the $l$-order streams are $\{s_{\mathcal{A}'}|\mathcal{A}'\subseteq\mathcal{K},|\mathcal{A}'|=l\}$ with $K\choose l$ distinct $l$-order streams included since there are $K\choose l$ different combinations of user subsets with cardinality equal to $l$. Let $\mathbf{s}_l\in\mathbb{C}^{{K\choose l}\times 1}$ denote the $l$-order data stream vector formed by all the $l$-order streams.  It is linearly precoded by the aggregate beamformer $\mathbf{P}_l\in \mathbb{C}^{ M\times {K\choose l}}$ formed by the beamforming vector of all the $l$-order streams, i.e., $\{\mathbf{p}_{\mathcal{A}'}|\mathcal{A}'\subseteq\mathcal{K},|\mathcal{A}'|=l\}$.   All the precoded streams are superposed into the transmit signal
\vspace{-2mm} 
\begin{equation}
\mathbf{{x}}=\mathbf{P}\mathbf{s}=\sum_{l=1}^{K}\mathbf{{P}}_{{l}}\mathbf{{s}}_{{l}}=\sum_{l=1}^{K}\sum_{\mathcal{A}'\subseteq\mathcal{K},|\mathcal{A}'|=l}\mathbf{{p}}_{\mathcal{A}'}{{s}}_{\mathcal{A}'}
\vspace{-1.5mm}
\end{equation}
 and broadcast to the users, where $\mathbf{s}=[\mathbf{s}_1^T,\ldots,\mathbf{s}_K^T]^T$ contains all the encoded streams and  $\mathbf{P}=[\mathbf{P}_1,\ldots,\mathbf{P}_K]$ is  the aggregate linear precoding matrix. Under the assumption that $\mathbb{E}\{\mathbf{{s}}\mathbf{{s}}^H\}=\mathbf{I}$,  the power constraint of each BS is $\left[\mathbf{P}\mathbf{P}^{H}\right]_{m,m}\leq P_{m},\forall m\in\mathcal{M}$. The signal received at each user is
$	y_{k}=\mathbf{{h}}_{k}^{H}\mathbf{{x}}+n_{k}, \forall k \in \mathcal{K}$,
where $\mathbf{h}_k=[h_{k,1},\ldots, h_{k,M}]^T\in\mathbb{C}^{M\times 1}$ is the aggregate channel from user-$k$ to  all  BSs and $h_{k,m}$ is the channel between BS-$m$ and user-$k$. $n_{k}\sim\mathcal{CN}(0,\sigma_{n,k}^2)$ is the Additive White Gaussian Noise (AWGN) received at user-$k$. The noise variance is assumed to be normalized without loss of generality at each user, i.e. $\sigma_{n,k}^2=1,\forall k\in\mathcal{K}$. The transmit SNR is equal to the total power consumption of all BSs $P_{tot}=\sum_{m\in\mathcal{M}}P_m$.

\par Each user decodes the streams that contains its intended sub-messages using SIC. The decoding order follows the rules that \textit{the data stream intended for more users has a higher decoding priority} \cite{RS2015bruno,Mingbo2016,RS2016hamdi,mao2017rate,mao2018rate,mao2018EE}. Therefore,  the streams with higher stream orders are always decoded before the streams with lower stream orders. The $K$-order stream is decoded first while the intended $1$-order stream is decoded last. As each user has multiple $l$-order streams when $1<l<K$, the decoding order of the intended streams with the same stream order is required to be optimized together with the beamformers.  Denote $\pi_l$ as one particular decoding order of all the $l$-order streams. The set of $l$-order streams to be decoded at user-$k$ is denoted as $\mathcal{S}_{l,k}=\{s_{\mathcal{A}'}|\mathcal{A}'\subseteq\mathcal{K},|\mathcal{A}'|=l,k\in\mathcal{A}'\}$. Sorted by the decoding order $\pi_l$, the elements of $\mathcal{S}_{l,k}$ form the $l$-order stream vector  $\mathbf{s}_{\pi_{l,k}}$, where $\mathbf{s}_{\pi_{l,k}}=[s_{\pi_{l,k}{(1)}},\cdots,s_{\pi_{l,k}{(|\mathcal{S}_{l,k}|)}}]^H$. $s_{\pi_{l,k}{(i)}}$ is assumed to be decoded before $s_{\pi_{l,k}{(j)}}$ if $i<j$. 
The Signal-to-Interference-plus-Noise Ratio (SINR) of user-$k$ to decode ${s}_{\pi_{l,k}{(i)}}$ is 
\vspace{-1.5mm}
\begin{equation}
\label{eq: sinr}
\small 
\gamma_{k}^{\pi_{l,k}{(i)}}=\frac{|\mathbf{{h}}_{k}^{H}\mathbf{{p}}_{\pi_{l,k}{(i)}}|^{2}}{I_{\pi_{l,k}{(i)}}+1},
\vspace{-1.5mm}
\end{equation}
where
\vspace{-1.5mm}
\begin{equation}
\label{eq: interference}
\begin{aligned}
	I_{\pi_{l,k}{(i)}}&=\sum_{j>i}|\mathbf{{h}}_{k}^{H}\mathbf{{p}}_{\pi_{l,k}(j)}|^{2}
	+\sum_{l'=1}^{l-1}\sum_{j=1}^{|\mathcal{S}_{l',k}|}|\mathbf{{h}}_{k}^{H}\mathbf{{p}}_{\pi_{l',k}(j)}|^{2}\\
	&+\sum_{\mathcal{A}'\subseteq\mathcal{K},k\notin\mathcal{A}'}|\mathbf{{h}}_{k}^{H}\mathbf{{p}}_{{\mathcal{A}'}}|^{2}
	\end{aligned}
	\vspace{-1.5mm}
\end{equation}
is the interference at user-$k$ to decode  ${s}_{\pi_{l,k}{(i)}}$. The first term  $\sum_{j>i}|\mathbf{{h}}_{k}^{H}\mathbf{{p}}_{\pi_{l,k}(j)}|^{2}$ is the interference from the $l$-order streams  going to be decoded after ${s}_{\pi_{l,k}{(i)}}$ in $\mathbf{s}_{{\pi_{l,k}}}$. The second term $\sum_{l'=1}^{l-1}\sum_{j=1}^{|\mathcal{S}_{l',k}|}|\mathbf{{h}}_{k}^{H}\mathbf{{p}}_{\pi_{l',k}(j)}|^{2}$ is the interference from lower order streams $\{\mathbf{s}_{{\pi_{l',k}}}| l'<l\}$ going to be decoded at user-$k$. The above interference is from the intended streams of user-$k$ while the third term $\sum_{\mathcal{A}'\subseteq\mathcal{K},k\notin\mathcal{A}'}|\mathbf{{h}}_{k}^{H}\mathbf{{p}}_{{\mathcal{A}'}}|^{2}$ is the interference from the streams that are not intended for user-$k$. These streams will not be decoded at user-$k$.
Based on (\ref{eq: sinr}), the achievable rate at user-$k$ to decode ${s}_{\pi_{l,k}{(i)}}$ is 
$
R_k^{\pi_{l,k}{(i)}}=\log_{2}(1+\gamma_{k}^{\pi_{l,k}{(i)}}).
$
The stream $s_{\mathcal{A}}$ will be decoded by all users in $\mathcal{A}$. To guarantee that the users in $\mathcal{A}$ can successfully decode  $s_{\mathcal{A}}$, the rate shall not exceed
$
R_{\mathcal{A}}=\min_{k'}\left\{ R_{k'}^{\mathcal{A}}\mid k'\in\mathcal{A}\right\}
$ \cite{RS2016hamdi}.
$R_{k}^{\mathcal{A}}$ is the rate of decoding the stream $s_{\mathcal{A}}$ at user-$k$ ($k\in\mathcal{A} $) based on the decoding order $\pi_{|\mathcal{A}|}$.  As $s_{\mathcal{A}}$ contains the sub-messages of users in $\mathcal{A}$, the rate $R_{k}^{\mathcal{A}}$ is shared by the users in $\mathcal{A}$.
Denote $C_k^{\mathcal{A}}$ as the portion of $R_{k}^{\mathcal{A}}$ allocated to user-$k$ $(k\in \mathcal{A})$ for the transmission of  sub-message $W_k^{\mathcal{A}}$, we have $\sum_{k'\in \mathcal{A}}C_{k'}^{\mathcal{A}}=R_{\mathcal{A}}$. The achievable rate of user-$k$ is 
$
R_{k,tot}=\sum_{\mathcal{A}'\subseteq\mathcal{K},k\in \mathcal{A}'}C_k^{\mathcal{A}'}+R_k.
$ $R_k$ is the  rate of decoding the 1-order stream $s_k$ at user-$k$.
By turning off the relevant messages, the above generalized RSMA framework  reduces to the existing SDMA, NOMA and RS models as discussed in \cite{mao2017rate}. This RSMA framework is more general and powerful than SDMA and NOMA and also leads to strategies with lower complexity than NOMA \cite{mao2017rate}. 
\begin{figure}[t!]
	\vspace{-1mm}
	\centering
	\includegraphics[width=3.2in]{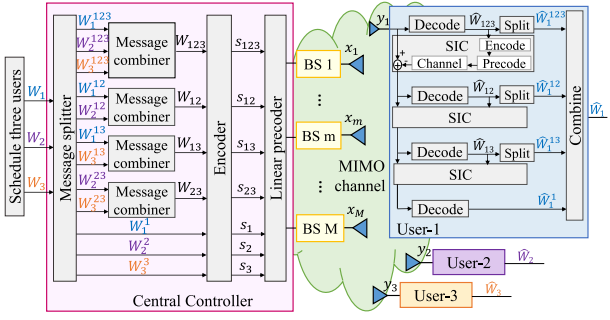} 
	\vspace{-0mm}
	\caption{Three-user RSMA-assisted CoMP JT model with $M$ BSs and decoding order $\pi_2= 12\rightarrow13\rightarrow23$.}
	\label{fig: three-user multi-cell}
	\vspace{-5mm}
\end{figure}

 \par To better illustrate the system model of RSMA in CoMP JT, we give an example of the $M$-cell three-user RSMA model in Fig. \ref{fig: three-user multi-cell}. At the central controller, the messages $W_1, W_2, W_3$ intended for the three users are respectively split into four sub-messages $\{W_1^{123}, W_1^{12}, W_1^{13}, W_1^{1}\}$, $\{W_2^{123}, W_2^{12}, W_2^{23}, W_2^{2}\}$, $\{W_3^{123}, W_3^{13}, W_3^{23}, W_3^{3}\}$. Sub-messages $\{W_1^{123},W_2^{123},W_3^{123}\}$ are jointly encoded into the 3-order stream $s_{123}$. The 2-order streams $s_{12}, s_{13}, s_{23}$ are respectively encoded by the sub-messages $\{W_1^{12}, W_2^{12}\}$, $\{W_1^{13}, W_3^{13}\}$, $\{W_2^{23}, W_3^{23}\}$. The encoded data streams are precoded and superposed to form the transmit signal $\mathbf{{x}}=\sum_{l=1}^{3}\mathbf{{P}}_{{l}}\mathbf{{s}}_{{l}}=\mathbf{{p}}_{{123}}{{s}}_{{123}}+\mathbf{{p}}_{{12}}{{s}}_{{12}}+\mathbf{{p}}_{{13}}{{s}}_{{13}}+\mathbf{{p}}_{{23}}{{s}}_{{23}}+\mathbf{{p}}_{{1}}{{s}}_{{1}}+\mathbf{{p}}_{{2}}{{s}}_{{2}}+\mathbf{{p}}_{{3}}{{s}}_{{3}}$, where $\mathbf{s}_1=[s_1,s_2,s_3]^T$, $\mathbf{s}_2=[s_{12},s_{13},s_{23}]^T$ and $\mathbf{s}_3=s_{123}$ are the formed 1-order, 2-order and 3-order data  stream vectors, respectively. The corresponding beamformers are $\mathbf{P}_1=[\mathbf{p}_1,\mathbf{p}_2,\mathbf{p}_3]$, $\mathbf{P}_2=[\mathbf{p}_{12},\mathbf{p}_{13},\mathbf{p}_{23}]$ and $\mathbf{P}_3=\mathbf{p}_{123}$.  Each element of the transmit signal $\mathbf{{x}}=[x_1,\ldots,x_m,\ldots,x_M]^T$ is the signal to be transmitted at the corresponding BS.  At user sides, each user decodes the streams that contains its intended sub-messages using SIC.  At user-1, the streams $s_{123}, s_{12}, s_{13}, s_1$ are decoded based on SIC.  Streams $s_{123}, s_{12}, s_{23}, s_2$ and $s_{123}, s_{13}, s_{23}, s_3$ are  decoded by user-2 and user-3, respectively. The 3-order stream $s_{123}$ is decoded first while the 1-order stream $s_k$ is decoded in the last place at user-$k$ ($\forall k\in\{1,2,3\}$). The decoding order $\pi_2$ of the 2-order streams $s_{12}, s_{13}, s_{23}$ is required to be optimized with the precoder.
In Fig. \ref{fig: three-user multi-cell}, we assume the decoding order of 2-order streams at all users are  $\pi_{2}= 12\rightarrow13\rightarrow23$.  Based on $\pi_{2}$, we specify the decoding procedure of user-1. The 2-order stream vectors to be decoded at user-1 is $\mathbf{s}_{\pi_{2,1}}=[s_{12},s_{13}]$.

The three-user RSMA illustrated above reduces to SDMA by simply allocating no power to the common streams $s_{123}, s_{12}, s_{13}, s_{23}$. Considering the other extreme of fully decoding the interference, there are two NOMA-assisted strategies, namely, `SC--SIC' and `SC--SIC per group' \cite{mao2017rate}. By allocating no power to the streams $s_1, s_2, s_{12}, s_{13}$, encoding $W_{1}$  into $s_{123}$, encoding $W_{2}$  into $s_{23}$ and encoding $W_{3}$  into $s_{3}$, the three-user RSMA reduces to SC--SIC with decoding order from the message of user-1 to that of user-3. By allocating no power to the streams $ s_{123}, s_{12},s_{13}$, encoding $W_{1}$  into $s_{1}$, encoding $W_{2}$  into $s_{23}$ and encoding $W_{3}$  into $s_{3}$, the three-user RSMA reduces to SC--SIC per group with user-1 in group-1 and user-2 and user-3 in group-2. The decoding order in group-2 is from the message of user-2 to that of user-3.

\vspace{-0.3mm}
\section{Problem Formulation and WMMSE Algorithm}
\vspace{-0.2mm}
\label{sec: problem formulation}

In this section, the  formulated  WSR maximization problem to design the precoder and the proposed  WMMSE-based optimization algorithm to solve the problem are specified.
\par We design the precoder of RSMA by studing the WSR maximization problem subject to individual QoS rate constraints and per-BS power constraint.  
For a given weight vector $\mathbf{u}=[u_1,\cdots,u_{K}]$ and a given stream decoding order  ${\pi}=\{\pi_1,\ldots,\pi_{K}\}$, the RSMA-based WSR maximization problem in  CoMP JT  is formulated as
\vspace{-1mm}
\begin{subequations}
\label{prob: K-user RS}
	\begin{align}
R_{\mathrm{RSMA}}(\mathbf{u},{\pi})&=\max_{\mathbf{{P}},\mathbf{c}}\sum_{k\in\mathcal{K}}u_{k}R_{k,tot} \label{o1}\\
\mbox{s.t.}\quad & \sum_{k'\in \mathcal{A}}C_{k'}^{\mathcal{A}}\leq R_{\mathcal{A}}, \forall \mathcal{A}\subseteq\mathcal{K} \label{c1}\\
&\left[\mathbf{P}\mathbf{P}^{H}\right]_{m,m}\leq P_m, \forall m\in\mathcal{M} \label{c2}\\
&   R_{k,tot}\geq R_k^{th}, \forall k\in\mathcal{K} \label{c3}\\
& \mathbf{c}\geq \mathbf{0}  \label{c4} 
\vspace{-2mm}
\end{align}
\end{subequations}
where $\mathbf{c}$ is the common rate vector formed by all the common rates $\{C_k^{\mathcal{A}}| \mathcal{A}\subseteq\mathcal{K}, k\in \mathcal{A}\}$. To maximize the WSR,  $\mathbf{c}$ and 
$\mathbf{P}$ should be jointly optimized. Constraint (\ref{c1}) is the common rate constraint. It ensures all the streams can be successfully decoded by the corresponding users. Constraint (\ref{c2})	is the power constraint of each BS. Constraint (\ref{c3}) is the QoS rate requirement of each user. $R_k^{th}$ is the lower bound of the achievable rate of user-$k$. Constraint (\ref{c4}) implies that the common rate of each stream allocated to each user is non-negative.
The rate region $R_{\mathrm{RSMA}}(\pi)$ of a certain decoding order $\pi$ is calculated by solving (\ref{prob: K-user RS}) using various $\mathbf{u}$.  The rate region of RSMA is the convex hull of the union over all decoding orders 
$
R_{\mathrm{RSMA}}=\mathrm{conv}\left(\bigcup_{\pi}R_{\mathrm{RSMA}}(\mathbf{\pi})\right).
$
\par Comparing the multi-cell Problem (\ref{prob: K-user RS})  with the single-cell problem (20) in \cite{mao2017rate}, the main difference lies in the power Constraint  (\ref{c2}). As each BS has its own power constraint, the sum power constraint in \cite{mao2017rate} is no longer applicable to the multi-cell case.   
By substituting the sum power constraint in the single-cell problem (20) of \cite{mao2017rate} with the per-BS power Constraint  (\ref{c2}), the WMMSE algorithm proposed in \cite{RS2016hamdi,mao2017rate} can be extended to solve (\ref{prob: K-user RS}). 

\par We introduce the variable $\mathbf{w}$ representing the Mean Square Error (MSE) weights formed by $\{{w}_k^{\mathcal{A}}| \mathcal{A}\subseteq\mathcal{K}, k\in \mathcal{A}\}$ and the  variable $\mathbf{g}$ representing the receive beamformer formed by $\{{g}_k^{\mathcal{A}}| \mathcal{A}\subseteq\mathcal{K}, k\in \mathcal{A}\}$, 
the original Problem (\ref{prob: K-user RS}) can be reformulated to the WMMSE problem, which is given by
\vspace{-2mm}
\begin{subequations}
	\label{prob: K-user WMMSE}
	\begin{align}
	&\min_{\mathbf{{P}},\mathbf{x},\mathbf{w},\mathbf{g}}\sum_{k\in\mathcal{K}}u_{k}\xi_{k,tot} \label{o1 wmmse}\\
	\mbox{s.t.}\quad & \sum_{k'\in \mathcal{A}}X_{k'}^{\mathcal{A}}+1\geq \xi_{\mathcal{A}}, \forall \mathcal{A}\subseteq\mathcal{K} \label{c1 wmmse}\\
	&\left[\mathbf{P}\mathbf{P}^{H}\right]_{m,m}\leq P_m, \forall m\in\mathcal{M} \label{c2 wmmse}\\
	&   \xi_{k,tot}\leq 1-R_k^{th}, \forall k\in\mathcal{K} \label{c3 wmmse}\\
	& \mathbf{x}\leq \mathbf{0}  \label{c4 wmmse} 
	\end{align}
\end{subequations}
where $\mathbf{x}$ is the transformation of the common rate vector $\mathbf{c}$, formed by $\{X_k^{\mathcal{A}}| \mathcal{A}\subseteq\mathcal{K}, k\in \mathcal{A}\}$. $\xi_{k,tot}$ is the total Weighted MSE (WMSE) formed by $\xi_{k,tot}=\sum_{\mathcal{A}'\subseteq\mathcal{K},k\in \mathcal{A}'}X_k^{\mathcal{A}'}+\xi_k^k$ and $\xi_{\mathcal{A}}=\max_{k'}\left\{ \xi_{k'}^{\mathcal{A}}\mid k'\in\mathcal{A}\right\}$. $\xi_k^{\mathcal{A}}$ is the augmented WMSE defined by 
$
	\xi_k^{\mathcal{A}}\triangleq w_k^{\mathcal{A}}\varepsilon_k^{\mathcal{A}}-\log_{2}(w_k^{\mathcal{A}}),
$
where $\varepsilon_k^{\mathcal{A}}$ is the MSE of decoding stream $s_{\mathcal{A}}$ at user-$k$ ($k\in\mathcal{A}$). Assuming that $s_{\mathcal{A}}$  is the $i$th stream to be decoded at user-$k$  according to the decoding order $\pi_{|\mathcal{A}|,k}$ of the $|\mathcal{A}|$-order streams, we have 
\vspace{-1mm}
\begin{equation}
	\varepsilon_k^{\mathcal{A}}\triangleq\mathbb{E}\{|\hat{s}_{\mathcal{A}}-s_{\mathcal{A}}|^{2}\}=g_k^{\mathcal{A}}T_k^{\mathcal{A}}-2\Re\{g_k^{\mathcal{A}}\mathbf{h}_k^H\mathbf{p}_{{\mathcal{A}}}\}+1,
	\vspace{-1mm}
\end{equation} where $T_k^{\mathcal{A}}\triangleq|\mathbf{h}_k^H\mathbf{P}_{\mathcal{A}}|^2+	I_{\pi_{|\mathcal{A}|,k}{(i)}}+1$. $I_{\pi_{|\mathcal{A}|,k}{(i)}}$ is calculated by Equation (\ref{eq: interference}). Readers are referred to Section 4.7 of \cite{mao2017rate} for the detailed rationale behind  the transformation from the original WSR problem to the WMMSE problem. 

\par The transformed WMMSE problem is still non-convex. However, it is block-wise convex in each block of $\mathbf{w}$, $\mathbf{g}$,  $(\mathbf{c}, \mathbf{P})$ when the other two blocks are fixed. When $\mathbf{g}$,  $(\mathbf{c}, \mathbf{P})$ are fixed, the optimal $\mathbf{w}^*=\mathbf{w}^{\textrm{MMSE}}$, where $\mathbf{w}^{\textrm{MMSE}}$ is formed by the MMSE weights $\{({w}_k^{\mathcal{A}})^{\textrm{MMSE}}| \mathcal{A}\subseteq\mathcal{K}, k\in \mathcal{A}\}$, and $({w}_k^{\mathcal{A}})^{\textrm{MMSE}}=(T_k^{\mathcal{A}}-|\mathbf{h}_k^H\mathbf{P}_{\mathcal{A}}|^2)^{-1}T_k^{\mathcal{A}}$.  When $\mathbf{w}$,  $(\mathbf{c}, \mathbf{P})$ are fixed, the optimal $\mathbf{g}^*=\mathbf{g}^{\textrm{MMSE}}$, where $\mathbf{g}^{\textrm{MMSE}}$ is formed by the MMSE equalizer $\{({g}_k^{\mathcal{A}})^{\textrm{MMSE}}| \mathcal{A}\subseteq\mathcal{K}, k\in \mathcal{A}\}$, and $({g}_k^{\mathcal{A}})^{\textrm{MMSE}}=\mathbf{p}_{\mathcal{A}}^H\mathbf{h}_k(T_k^{\mathcal{A}})^{-1}$. When $ \mathbf{u}, \mathbf{g}$ are fixed, $(\mathbf{x}, \mathbf{P})$ are coupled in (\ref{prob: K-user WMMSE}). Problem (\ref{prob: K-user WMMSE}) becomes a convex Quadratically Constrained Quadratic Program
(QCQP) and can be solved using the interior-point
method.  It can be easily shown that any solution $(\mathbf{c}^*, \mathbf{P}^*)$ satisfying the KKT optimality conditions of the WSR maximization Problem (\ref{prob: K-user RS}) is also a  solution $(\mathbf{x}^*, \mathbf{P}^*)$ satisfying the KKT optimality conditions of  the  WMMSE Problem (\ref{prob: K-user WMMSE}) with $\mathbf{x}^*=-\mathbf{c}^*$, $\mathbf{w}^*=\mathbf{w}^{\textrm{MMSE}}$, $\mathbf{g}^*=\mathbf{g}^{\textrm{MMSE}}$  holds at the optimal solution.
These properties motivates us to use AO algorithm to
solve the problem, as shown in Algorithm 1.  In each iteration $n$, the MSE weights $\mathbf{w}$ and receive beamformer $\mathbf{g}$ are updated based on the precoder in the previous iteration $\mathbf{P}^{[n-1]}$. $(\mathbf{x}, \mathbf{P})$ are then updated by solving Problem (\ref{prob: K-user WMMSE}). The variables are iteratively updated until the WSR converges. As the WSR is increasing  iteratively and the problem is bounded above for given per-BS power constraints, the proposed AO algorithm is guaranteed to converge.

\setlength{\textfloatsep}{5pt}	
\begin{algorithm}[t!]
	\textbf{Initialize}: $n\leftarrow0$, $\mathbf{P}^{[n]}$, $\mathrm{WSR}^{[n]}$\;
	\Repeat{$|\mathrm{WSR}^{[n]}-\mathrm{WSR}^{[n-1]}|\leq \epsilon$}{
		$n\leftarrow n+1$\;
		$\mathbf{P}^{[n-1]}\leftarrow \mathbf{P}$\;
		$\mathbf{w}\leftarrow\mathbf{w}^{\mathrm{MMSE}}(\mathbf{P}^{[n-1]})$; $\mathbf{g}\leftarrow\mathbf{g}^{\mathrm{MMSE}}(\mathbf{P}^{[n-1]})$\;
		update $(\mathbf{x},\mathbf{P})$ by solving (\ref{prob: K-user WMMSE}) using the updated $\mathbf{w}, \mathbf{g}$;	
	}	
	\caption{WMMSE-based AO algorithm}
	\label{WMMSE algorithm}					
\end{algorithm}

%

\vspace{-1mm}
\section{Numerical Results}
\vspace{-0.5mm}
\label{sec: simulation}
In this section, the performance of the proposed RSMA in a CoMP JT network is evaluated.  We first study the case of two cells cooperatively serving two users followed by the three-cell three-user case\footnote{We consider cooperation of two and three cells because these are the typical numbers of JT. It is known that there is little benefit in considering cooperation among a larger number of cells in homogeneous networks \cite{clerckx2013mimo}.}. We assume the system is abstracted from the Wyner model \cite{gesbert2010multicell}. The cells are arranged in a linear array and interference only comes from immediate neighboring cells. The channel $h_{k,m}$ from BS in cell-$m$ to user-$k$ is assumed to be  independent and identically distributed (i.i.d) complex Gaussian with zero mean and variance $\sigma_{k,m}^2\triangleq\mathbb{E}[|h_{k,m}|^2]$, i.e., $h_{k,m}\sim\mathcal{CN}(0,\sigma_{k,m}^2),\forall k\in\mathcal{K}, m\in\mathcal{M}$. 
\vspace{-2mm}
\subsection{Two-cell case}
\vspace{-1mm}
We first consider a two-cell cooperative transmission system ($M=2$) with one user in each cell ($K_m=1$). The channels of user-1 and user-2 are 
$
\mathbf{h}_1=[h_{1,1},h_{1,2}]^T$ and
$\mathbf{h}_2=[h_{2,1},h_{2,2}]^T
$, respectively. The variances of the channels are varied as $\sigma_{1,1}^2=1$, $\sigma_{1,2}^2=\alpha$, $\sigma_{2,1}^2=\alpha\beta$, $\sigma_{2,2}^2=\beta$.  $\alpha\in(0,1]$ represents the disparity of channel strengths across the BSs. 
When $\alpha=1$, the channel strengths from both BSs to a given user are the same.  
As $\alpha$ decreases, the channel strength from the adjacent BS decreases while both users experience stronger channel strengths from their corresponding serving BSs.  When $\alpha=0$, each user is served by the serving BS only. Note that, as $\alpha$ decreases, the channels of the users become more orthogonal to each other.  $\beta\in(0,1]$ represents the disparity of channel strengths between the users. When $\beta=1$, the channel strengths from the BSs to both users, measured here in terms of average channel vector norm, are the same. As $\beta$ decreases, the disparity of channel strengths  between the users increases. User-2 suffers a more severe path loss.

The two-user rate region are illustrated and the boundary of the rate region is obtained by solving Problem (\ref{prob: K-user RS}) with various weights $\mathbf{u}$. The weights chosen in this work follows \cite{mao2017rate}. $u_1$ is fixed to 1 while $u_2$ is varied as $u_2=10^{[-3,-1,-0.95,\ldots,0.95,1,3]}$. To find the largest rate region, the QoS rate requirements are set to 0, i.e., ${R}_{th}=0 $ bit/s/Hz.  The rate region of RSMA is compared with that of SDMA based on MU--LP and NOMA based on SC--SIC specified in \cite{mao2017rate}. To simplify the notation, RS, MU--LP and SC--SIC are used to represent RSMA, SDMA  and NOMA, respectively. As the capacity region of Multiple Input Single Output (MISO) BC with PAPC is achieved by DPC, we compare with the DPC region generated by the AO algorithm in \cite{pham2017alternating}.  As mentioned in Section \ref{sec: system model}, $\sigma_{n,k}^2=1,\forall k\in\mathcal{K}$. By setting the transmit SNR to 20 dB, the total power constraint across all BSs  becomes $P_{tot}=100$ Watt. The power limit of each BS is chosen as $P_m=\frac{P_{tot}}{M}$. The beamformer initialization of the WMMSE algorithm follows \cite{mao2017rate}.

\begin{figure}[t!]
	\vspace{-5mm}
	\centering
	\includegraphics[width=2.8in]{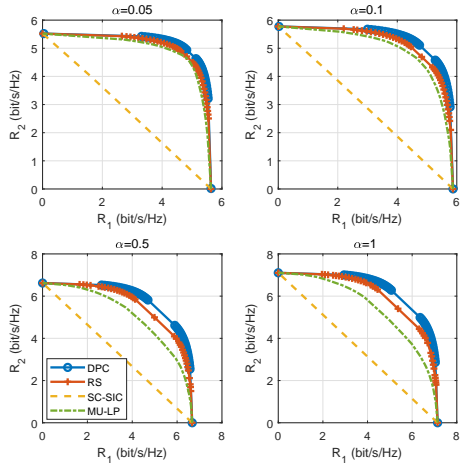}
	\vspace{-3mm}
	\caption{Achievable rate region comparison of different strategies, averaged over 100 random channel realizations, $\beta=1$.} 
	\label{fig: rate region SNR 20dB beta1}
	\vspace{-4mm}
\end{figure}
\begin{figure}[t!]
	\vspace{-2mm}
	\centering
	\includegraphics[width=2.8in]{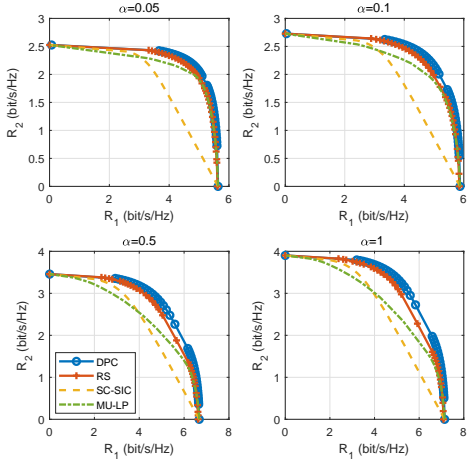}
	\vspace{-3mm}
	\caption{Achievable rate region comparison of different strategies, averaged over 100 random channel realizations, $\beta=0.1$.}
	\label{fig: rate region SNR 20dB beta01}
	\vspace{-0mm}
\end{figure}
Fig. \ref{fig: rate region SNR 20dB beta1} and Fig. \ref{fig: rate region SNR 20dB beta01} show the average rate region comparison of different strategies over 100 random channel realizations with   $\beta=1$ and $\beta=0.1$, respectively.  For each $\beta$, the results of $\alpha=[0.05, 0.1, 0.5, 1]$ are illustrated. In each figure, the rate regions of all strategies increase with $\alpha$. Comparing the corresponding subfigures of Fig. \ref{fig: rate region SNR 20dB beta01} and Fig. \ref{fig: rate region SNR 20dB beta1}, the rate regions of all strategies increase with $\beta$. 
The cooperation of the BSs effectively forms a MISO BC. As $\alpha$ and $\beta$ increase, the user channel gains increase. Therefore, the rate performance is improved with $\alpha$ and $\beta$. 
In all subfigures, the rate region of RS is larger than that of SC--SIC and MU--LP. It is closer to the capacity region achieved by DPC.
SC--SIC has the worst performance in Fig. \ref{fig: rate region SNR 20dB beta1}.  As the SC--SIC scheme is motivated by leveraging the disparity of channel strengths among users, it is not suitable when there is no channel strength disparity. When $\alpha$ decreases, the rate region gap between RS and MU--LP decreases.  This is because the channels of the two users become more orthogonal with each other as $\alpha$ decreases, making it more suitable for MU--LP.
Comparing Fig. \ref{fig: rate region SNR 20dB beta01} and Fig. \ref{fig: rate region SNR 20dB beta1}, the rate region gap between RS and SC--SIC decreases while the rate region gap between RS and MU--LP increases as $\beta$ decreases. When $\beta=0.1$, SC--SIC works better due to the disparity of channel strengths among users. However, the performance of SC--SIC becomes worse as $\alpha$ decreases. SC--SIC is not suitable when the channels of the users are (semi-) orthogonal. In contrast, MU--LP leverages the orthogonality of channels and gets closer to RS. In all subfigures of Fig. \ref{fig: rate region SNR 20dB beta01}, SC--SIC and MU--LP outperform each other at one part of the rate region. The rate region of RS is larger than the convex hull of the rate regions of SC--SIC and MU--LP. 
In summary, SDMA performs better for large $\beta$  and small $\alpha$ (similar strengths and orthogonal) while NOMA performs better for small $\beta$  and large $\alpha$ (channel closer to alignment and disparity in channel strengths). RS is more robust as it copes with all deployments (for all  $\beta$ and $\alpha$).




\vspace{-1.0mm}
\subsection{Three-cell case}
\vspace{-1.0mm}
\begin{figure}[t!]
	\vspace{-2mm}
	\centering
	\includegraphics[width=2in]{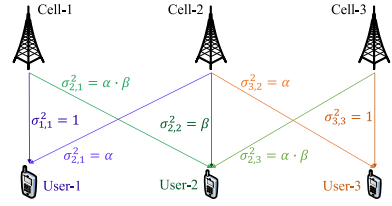}
	\vspace{-1mm}
	\caption{Three-user Wyner model, $M=3, K=3$.}
	\label{fig: wyner 3}
	\vspace{-1mm}
\end{figure}
We further consider a three-cell cooperative transmission system ($M=3$) and there is one user in each cell ($K_m=1$). There are in total $K=3$ users. The three-user Wyner model is shown in Fig. \ref{fig: wyner 3}. Based on the Wyner model, the channels of user-1 to user-3 are given by 
$
\mathbf{h}_1=[h_{11},h_{12},0]^T, 
\mathbf{h}_2=[h_{21},h_{22},h_{23}]^T,
\mathbf{h}_3=[0,h_{32},h_{33}]^T.
$
The variances of the channels are $\sigma_{1,1}^2=1$, $\sigma_{1,2}^2=\alpha$, $\sigma_{2,1}^2=\sigma_{2,3}^2=\alpha\beta$, $\sigma_{2,2}^2=\beta$, $\sigma_{3,2}^2=\alpha$, $\sigma_{3,3}^2=1$.  $\beta\in(0,1]$ represents the disparity of channel strengths between user-1 and user-2 as well as user-2 and user-3.
To further investigate the performance of RSMA, we also compare the performance of RSMA with 1-layer RS and SC--SIC per group as specified in \cite{mao2017rate}. The 1-layer RS is a low-complexity sub-scheme of the generalized RS framework with only the $K$-order common stream and 1-order private streams to be transmitted. Each user first decodes the common stream and then its intended private stream. Only one SIC is required at each user.  SC--SIC per group separates users into multiple groups and the users within each group are served using SC--SIC while the users across the groups are served using SDMA. The complexity of SC--SIC per group increases as the user ordering and grouping is required to be optimized jointly. In this work, we consider a fixed grouping method where user-1 is in group-1, user-2 and user-3 are in group-2. The decoding order is optimized with the precoder \cite{mao2017rate}.

\begin{figure}[t!]
	\vspace{-3mm}
	\centering
	\includegraphics[width=2.9in]{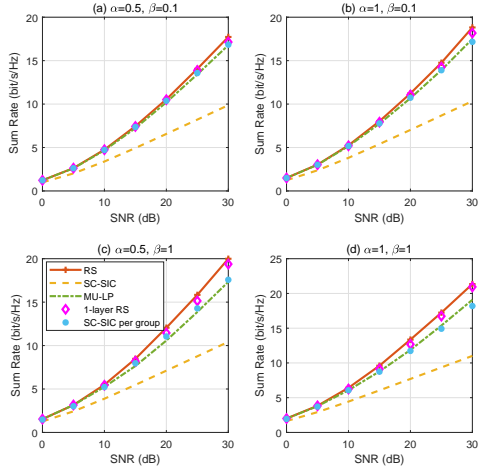}
	\vspace{-3mm}
	\caption{Sum rate versus SNR comparison of different strategies, averaged over 10 random channel realizations,  $u_1=1, u_2=1, u_3=1$, $\mathbf{r}_{th}=[0.001,0.01,0.03,0.08,0.1,0.1,0.1]$ bit/s/Hz.}
	\label{fig: sum rate vs snr}
	\vspace{-2mm}
\end{figure}

Fig. \ref{fig: sum rate vs snr} shows the  sum rate versus SNR comparisons of different strategies. $\alpha,\beta$ are varied in each subfigure.  The rate threshold is changing as $\mathbf{r}_{th}=[0.001,0.01,0.03,0.08,0.1,0.1,0.1]$ bit/s/Hz for $\mathrm{SNR}=[0,5,10,15,20,25,30]$ dBs.  In all subfigures, RS and 1-layer RS show clear sum rate improvements over all existing schemes. Only one layer of SIC is required at each receiver in 1-layer RS. In this setup, 1-layer RS has the same receiver complexity as SC--SIC per group and a lower receiver complexity than SC--SIC, and additionally does not require any grouping and ordering optimization.   In contrast, SC--SIC has the worst performance due to the loss of multiplexing gain.  SC--SIC forces one receiver to decode messages of all other users which leads to a collapse of the sum-Degrees-of-Freedom (DoF) to 1. Therefore, the multiplexing gain is lost. SC--SIC per group outperforms SC--SIC in all subfigures as SC--SIC per group treats inter-group interference as noise\footnote{Note that this is not a general observation and there are instances where SC-SIC outperforms SC-SIC per group as shown in \cite{mao2017rate}.}. 
In all subfigures, the sum rate of all the schemes are increasing with $\alpha,\beta$.  The channel strength becomes stronger as $\alpha,\beta$ increase, resulting in the increase of the system sum rate.

\begin{figure}[t!]
	\vspace{-1mm}
	\centering
	\includegraphics[width=2.9in]{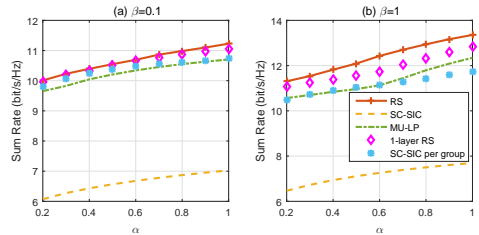}
	\vspace{-2mm}
	\caption{Sum rate versus $\alpha$ comparison of different strategies, averaged over 10 random channel realizations,  SNR=20 dB, $u_1=1, u_2=1, u_3=1$, ${R}_{th}=0.1$ bit/s/Hz.}
	\label{fig: sum rate vs alpha SNR 20 dB}
	\vspace{-0mm}
\end{figure}

Fig. \ref{fig: sum rate vs alpha SNR 20 dB} shows the  sum rate versus $\alpha$ for different schemes. In each subfigure, the sum rate of each scheme is increasing with $\alpha$. There is a significant sum rate improvement of RS-based schemes over all existing schemes.  Thanks to their abilities of partially decoding the interference and partially treating the interference as noise, RS-based schemes overcome the limitations of existing schemes by dynamically adjusting the amount of interference  treated as noise (through the presence of 1-order streams) and decoded by the users (through the presence of higher-order streams)  to the
channel conditions.

	\vspace{-1.5mm}
\section{Conclusions}
\vspace{-1mm}
\label{sec: conclusion}
To conclude, we initiate the investigation of RSMA in CoMP JT. The WSR maximization problem with QoS rate constraints and per-BS power constraints is solved using the WMMSE algorithm. We show in the numerical results that SDMA is more suited to the scenarios where there is little inter-user channel strength disparity but large inter-cell channel disparity. NOMA is more suited to the scenarios where there is a large inter-user channel strength disparity but little inter-cell channel disparity.  In comparison, RSMA always bridges, generalizes and outperforms existing SDMA and NOMA strategies. It is suited to any deployment with any inter-user and inter-cell channel disparities. 1-layer RS shows great benefit of reducing the transmitter and receiver complexity while maintaining better performance than SDMA and NOMA strategies.
Therefore, RSMA is a more general, robust and powerful multiple access scheme for downlink CoMP JT networks.

\bibliographystyle{IEEEtran}
\vspace{-2mm}
\bibliography{reference}	
\vspace{-4mm}

\end{document}